\documentstyle[twoside,epsfig]{article}

\catcode`\@=11
\long\def\@makefntext#1{
\protect\noindent \hbox to 3.2pt {\hskip-.9pt
$^{{\eightrm\@thefnmark}}$\hfil}#1\hfill}       

\def\@makefnmark{\hbox to 0pt{$^{\@thefnmark}$\hss}}    

\def\ps@myheadings{\let\@mkboth\@gobbletwo
\def\@oddhead{\hbox{}
\rightmark\hfil\eightrm\thepage}
\def\@oddfoot{}\def\@evenhead{\eightrm\thepage\hfil
\leftmark\hbox{}}\def\@evenfoot{}
\def\sectionmark##1{}\def\subsectionmark##1{}}

\oddsidemargin=\evensidemargin
\newcounter{sectionc}\newcounter{subsectionc}\newcounter{subsubsectionc}
\renewcommand{\section}[1] {\vspace{12pt}\addtocounter{sectionc}{1}
\setcounter{subsectionc}{0}\setcounter{subsubsectionc}{0}\noindent
    {\tenbf\thesectionc. #1}\par\vspace{5pt}}
\renewcommand{\subsection}[1] {\vspace{12pt}\addtocounter{subsectionc}{1}
    \setcounter{subsubsectionc}{0}\noindent
    {\bf\thesectionc.\thesubsectionc. {\kern1pt \bfit #1}}\par\vspace{5pt}}
\renewcommand{\subsubsection}[1] {\vspace{12pt}\addtocounter{subsubsectionc}{1}
    \noindent{\tenrm\thesectionc.\thesubsectionc.\thesubsubsectionc.
    {\kern1pt \tenit #1}}\par\vspace{5pt}}
\newcommand{\nonumsection}[1] {\vspace{12pt}\noindent{\tenbf #1}
    \par\vspace{5pt}}

\newcounter{appendixc}
\newcounter{subappendixc}[appendixc]
\newcounter{subsubappendixc}[subappendixc]
\renewcommand{\thesubappendixc}{\Alph{appendixc}.\arabic{subappendixc}}
\renewcommand{\thesubsubappendixc}
    {\Alph{appendixc}.\arabic{subappendixc}.\arabic{subsubappendixc}}

\renewcommand{\appendix}[1] {\vspace{12pt}
        \refstepcounter{appendixc}
        \setcounter{figure}{0}
        \setcounter{table}{0}
        \setcounter{lemma}{0}
        \setcounter{theorem}{0}
        \setcounter{corollary}{0}
        \setcounter{definition}{0}
        \setcounter{equation}{0}
        \renewcommand{\thefigure}{\Alph{appendixc}.\arabic{figure}}
        \renewcommand{\thetable}{\Alph{appendixc}.\arabic{table}}
        \renewcommand{\theappendixc}{\Alph{appendixc}}
        \renewcommand{\thelemma}{\Alph{appendixc}.\arabic{lemma}}
        \renewcommand{\thetheorem}{\Alph{appendixc}.\arabic{theorem}}
        \renewcommand{\thedefinition}{\Alph{appendixc}.\arabic{definition}}
        \renewcommand{\thecorollary}{\Alph{appendixc}.\arabic{corollary}}
        \renewcommand{\theequation}{\Alph{appendixc}.\arabic{equation}}
        \noindent{\tenbf Appendix \theappendixc #1}\par\vspace{5pt}}
\newcommand{\subappendix}[1] {\vspace{12pt}
        \refstepcounter{subappendixc}
        \noindent{\bf Appendix \thesubappendixc. {\kern1pt \bfit #1}}
    \par\vspace{5pt}}
\newcommand{\subsubappendix}[1] {\vspace{12pt}
        \refstepcounter{subsubappendixc}
        \noindent{\rm Appendix \thesubsubappendixc. {\kern1pt \tenit #1}}
    \par\vspace{5pt}}

\newcommand{\textlineskip}{\baselineskip=13pt}
\newcommand{\smalllineskip}{\baselineskip=10pt}

\def\eightcirc{
\begin{picture}(0,0)
\put(4.4,1.8){\circle{6.5}}
\end{picture}}
\def\eightcopyright{\eightcirc\kern2.7pt\hbox{\eightrm c}}

\newcommand{\copyrightheading}[1]
    {\vspace*{-2.5cm}\smalllineskip{\flushleft
    {\footnotesize Modern Physics Letters A #1}\\
    {\footnotesize Vol. 18, Nos. 2-6 (2003) 248-257 #1}\\
    {\footnotesize $\eightcopyright$\, World Scientific Publishing
     Company}\\
     }}

\def\abstracts#1#2#3{{
    \centering{\begin{minipage}{4.5in}\footnotesize\baselineskip=10pt
    \parindent=0pt #1\par
    \parindent=15pt #2\par
    \parindent=15pt #3
    \end{minipage}}\par}}

\newcommand{\bibit}{\nineit}
\newcommand{\bibbf}{\ninebf}
\renewenvironment{thebibliography}[1]
    {\frenchspacing
     \ninerm\baselineskip=11pt
     \begin{list}{\arabic{enumi}.}
        {\usecounter{enumi}\setlength{\parsep}{0pt}
     \setlength{\leftmargin 12.7pt}{\rightmargin 0pt} 
         \setlength{\itemsep}{0pt} \settowidth
    {\labelwidth}{#1.}\sloppy}}{\end{list}}

\newcounter{itemlistc}
\newcounter{romanlistc}
\newcounter{alphlistc}
\newcounter{arabiclistc}

\newcommand{\fcaption}[1]{
        \refstepcounter{figure}
        \setbox\@tempboxa = \hbox{\footnotesize Fig.~\thefigure. #1}
        \ifdim \wd\@tempboxa > 5in
           {\begin{center}
        \parbox{5in}{\footnotesize\smalllineskip Fig.~\thefigure. #1}
            \end{center}}
        \else
             {\begin{center}
             {\footnotesize Fig.~\thefigure. #1}
              \end{center}}
        \fi}

\newcommand{\tcaption}[1]{
        \refstepcounter{table}
        \setbox\@tempboxa = \hbox{\footnotesize Table~\thetable. #1}
        \ifdim \wd\@tempboxa > 5in
           {\begin{center}
        \parbox{5in}{\footnotesize\smalllineskip Table~\thetable. #1}
            \end{center}}
        \else
             {\begin{center}
             {\footnotesize Table~\thetable. #1}
              \end{center}}
        \fi}

\def\@citex[#1]#2{\if@filesw\immediate\write\@auxout
    {\string\citation{#2}}\fi
\def\@citea{}\@cite{\@for\@citeb:=#2\do
    {\@citea\def\@citea{,}\@ifundefined
    {b@\@citeb}{{\bf ?}\@warning
    {Citation `\@citeb' on page \thepage \space undefined}}
    {\csname b@\@citeb\endcsname}}}{#1}}

\newif\if@cghi
\def\cite{\@cghitrue\@ifnextchar [{\@tempswatrue
    \@citex}{\@tempswafalse\@citex[]}}
\def\citelow{\@cghifalse\@ifnextchar [{\@tempswatrue
    \@citex}{\@tempswafalse\@citex[]}}
\def\@cite#1#2{{$\null^{#1}$\if@tempswa\typeout
    {IJCGA warning: optional citation argument
    ignored: `#2'} \fi}}

\def\pmb#1{\setbox0=\hbox{#1}
    \kern-.025em\copy0\kern-\wd0
    \kern.05em\copy0\kern-\wd0
    \kern-.025em\raise.0433em\box0}

\def\fnt#1#2{\footnotetext{\kern-.3em
    {$^{\mbox{\scriptsize #1}}$}{#2}}}

\def\ps@myheadings{%
    \let\@oddfoot\@empty\let\@evenfoot\@empty
    \def\@evenhead{\slshape\leftmark\hfil}
    \def\@oddhead{\hfil{\slshape\rightmark}}
    \let\@mkboth\@gobbletwo
    \let\sectionmark\@gobble
    \let\subsectionmark\@gobble
    }
\font\tenrm=cmr10
\font\tenit=cmti10
\font\tenbf=cmbx10
\font\bfit=cmbxti10 at 10pt
\font\ninerm=cmr9
\font\nineit=cmti9
\font\ninebf=cmbx9
\font\eightrm=cmr8

\def\qed{\hbox{${\vcenter{\vbox{            
   \hrule height 0.4pt\hbox{\vrule width 0.4pt height 6pt
   \kern5pt\vrule width 0.4pt}\hrule height 0.4pt}}}$}}

\begin{document}
\setlength{\textheight}{7.7truein}  

\thispagestyle{empty}


\markboth{\protect{\footnotesize\it Probing $\Delta$ structure...}}
{\protect{\footnotesize\it Probing $\Delta$ structure...}}
\normalsize\textlineskip

\setcounter{page}{1}

\copyrightheading{} 

\vspace*{0.88truein}

\centerline{\bf  PROBING $\Delta$ STRUCTURE }
\baselineskip=13pt
\centerline{\bf WITH PION ELECTROMAGNETIC PRODUCTION}
\vspace*{0.4truein}
\centerline{\footnotesize SHIN NAN YANG}
\baselineskip=12pt
\centerline{\footnotesize\it Physics Department, National Taiwan University
}
\baselineskip=10pt
\centerline{\footnotesize\it Taipei 10617,  Taiwan
}
\vspace*{12pt}

\centerline{\footnotesize  SABIT S. KAMALOV}
\baselineskip=12pt
\centerline{\footnotesize\it Bogoliubov Laboratory for Theoretical Physics, JINR}
\baselineskip=10pt
\centerline{\footnotesize\it Dubna, 141980 Moscow Region, Russia}
\vspace*{0.228truein}


\vspace*{0.23truein}
\abstracts{The Dubna-Mainz-Taipei dynamical model for pion electromagnetic production,
which can describe well the existing data from threshold up to 1 GeV photon
lab energy,
is presented and used to analyze the recent precision data in the $\Delta$ region.
We find that, within our model, the bare $\Delta$ is almost spherical while
 the physical $\Delta$ is oblate. The deformation is almost
saturated by the pion cloud effects. We further find that up to $Q^2 = 4.0
(GeV/c)^2$, the extracted helicity amplitude $A_{3/2}$ and $A_{1/2}$ remain
comparable with each other, implying that hadronic helicity is not conserved
at this range of $Q^2$. The ratio $E_{1+}/M_{1^+}$ obtained show, starting 
from a small and negative value at the real photon point, a clear tendency
to cross zero, and to become positive with increasing $Q^2$. This is a possible
indication of a very slow approach toward the pQCD region. Finally, we find
that the bare helicity amplitude $A_{1/2}$ and $S_{1/2}$, but not $A_{3/2}$,
starts exhibiting the scaling behavior at about $Q^2 \ge 2.5 (GeV/c)^2$.}{}{}


\vspace*{2pt}


\baselineskip=13pt          
\normalsize                 


\section{Introduction}      
\vspace*{-0.5pt} \noindent
$\Delta$ is the first excited state of
the nucleon and the only well isolated nucleon resonance. Its
properties serve as a bench mark for models of nucleon structure.
It is hence important to extract $\Delta$'s properties from
experiments reliably.

There are two kinds of electromagnetic properties of the $\Delta$.
The first involves the $\Delta$ itself, like the magnetic dipole
moment $\mu_\Delta$ and electric quadrupole moment $Q_\Delta$.
They are difficult to measure because of the short life time of
the $\Delta$. The others involve the $N \rightarrow \Delta$
transition, like $\mu_{N\rightarrow\Delta}$ and
$Q_{N\rightarrow\Delta}.$ They can be determined from the pion
electromagnetic production via the following relations,
\begin{eqnarray}
\mu_{N\rightarrow\Delta} &=& - \frac {m_N}{2\sqrt{\pi\alpha_e\omega}}
                              (A_{1/2}^{\Delta}+\sqrt{3}A_{3/2}^{\Delta}), \label{eq:muDelta}\\
Q_{N\rightarrow\Delta}   &=& -\frac {3}{2\sqrt{\pi\alpha_e\omega^3}}
                              (A_{1/2}^{\Delta}-\frac{1}{\sqrt{3}}A_{3/2}^{\Delta}),\label{eq:QDelta}
\end{eqnarray}
where $A's$ are the helicity amplitudes and $\omega$ the photon energy.

In this talk, we'll focus mostly on the $\Delta$ properties
associated with the $\gamma^* N \leftrightarrow \Delta$
transition. They are of interest because in symmetric $SU(6)$
quark models and with the inclusion of only one-body current
contribution \cite{buchmann01}, the $\gamma N\Delta$ transition
can proceed only via the flip of a single quark spin in the
nucleon, leading  to $M_{1+}$ dominance and $E_{1+} = S_{1+}
\equiv 0$. If the $\Delta$ is deformed, then the photon can excite
a nucleon into a $\Delta$ through electric quadrupole E2 and
Coulomb quadrupole C2 transitions. Recent experiments give
nonvanishing ratio $R_{EM}
=E_{1+}^{(3/2)}/M_{1+}^{(3/2)}$ lying between $-2.5\%$~\cite{Beck97}
and $-3.0\%$~\cite{Blanpied97}, or equivalently
$Q_{N\rightarrow\Delta} \simeq -0.108 \,fm^2$, at $Q^2=0$. This
has been widely taken as an indication of  a deformed (oblate)
$\Delta$, namely, an admixture of a D state in the $\Delta$. On
the other hand, in the limit of $Q^2 \rightarrow \infty$, pQCD
predicts the dominance of helicity-conserving amplitudes
\cite{Brodsky81} and scaling results \cite{Carlson,Stoler}. The
hadronic helicity conservation should have the consequence that
$R_{EM}$ approaches 1. The scaling
behavior predicted by pQCD for the helicity amplitudes is
$A_{1/2}^{\Delta}\sim Q^{-3}$, $A_{3/2}^{\Delta}\sim Q^{-5}$, and
the Coulomb helicity amplitude $S_{1/2}^{\Delta}\sim Q^{-3}$,
resulting in $R_{SM} =S_{1+}^{(3/2)}/M_{1+}^{(3/2)} \rightarrow
const$. Accordingly, the question of how $R_{EM}$ would evolve
from a very small negative value at $Q^2 = 0$ to $+100\%$ at
sufficiently high $Q^2$, has attracted  great interest both
theoretically and experimentally.

Because of the significance of the physics involved in the $Q^2$
evolution of $R_{EM}$ and $R_{SM}$, it is important to employ the
best possible extraction method in the analysis of the data. So we
now turn to the theoretical method in describing the pion
electromagnetic production.

\vspace*{2pt}
\section{Dynamical Model for Pion Electromagnetic Production}
\noindent
At present, there exist three different theoretical
methods to describe the pion electromagnetic production. The
oldest one is the dispersion theory which was developed by Chew
{\it et al.} \cite{CGLN57}, and is still in use today \cite{HDT}.
It is based on analyticity, unitarity and crossing symmetry. The
second method was based on chiral Lagrangians as carried out by
Olsson and Osypowski \cite{Olsson75}. This work was further
developed by Davidson {\it et al.} \cite{Davidson91}. MAID
\cite{MAID} is also constructed along this line. In 1985, Yang
\cite{Yang85} and Tanabe and Ohta \cite{Tanabe85} developed the
dynamical model of pion photoproduction which has been currently
in wide use \cite{NBL90,Surya96,SL96,Yang96}. We have recently
constructed a DMT dynamical model \cite{KY99,Kamalov01} which can
describe the existing pion production data from threshold
\cite{Kamalov01a} to $1\, GeV$ photon lab energy. We will use DMT
model for our analysis hereafter.

In the dynamical approach to pion photo- and electroproduction
\cite{Yang85}, the t-matrix can be expressed as
\begin{eqnarray}
t_{\gamma\pi}(E)=v_{\gamma\pi}+v_{\gamma\pi}\,g_0(E)\,t_{\pi
N}(E), \label{eq:tgammapi}
\end{eqnarray}
and the physical multipoles in channel $\alpha$ are given by
\begin{eqnarray}
& &t_{\gamma\pi}^{(\alpha)}(q_E,k;E+i\epsilon)
=\exp{(i\delta^{(\alpha)})}\,\cos{\delta^{(\alpha)}}
\nonumber\\&\times& \left[v_{\gamma\pi}^{(\alpha)}(q_E,k) +
P\int_0^{\infty} dq' \frac{q'^2R_{\pi
N}^{(\alpha)}(q_E,q')\,v_{\gamma\pi}^{(\alpha)}(q',k)}{E-E_{\pi
N}(q')}\right], \label{eq:backgr}
\end{eqnarray}
where $v_{\gamma\pi}$ is the transition potential for $\gamma^*N
\rightarrow \pi N$, and $t_{\pi N}$ and $g_0$ denote the $\pi N$
scattering t-matrix and free propagator, respectively, with $E
\equiv W$ the total energy in the CM frame. $\delta^{(\alpha)}$
and $R_{\pi N}^{(\alpha)}$ are the $\pi N$ scattering phase shift
and reaction matrix in channel $\alpha$, respectively; $q_E$ is
the pion on-shell momentum and $k=|{\bf k}|$ is the photon
momentum.

In a resonant channel like (3,3) in which the $\Delta(1232)$ plays
a dominant role, the transition potential $v_{\gamma\pi}$ consists
of two terms
\begin{eqnarray}
v_{\gamma\pi}(E)=v_{\gamma\pi}^B +v_{\gamma\pi}^{\Delta}(E),\label{eq:vgammapi33}
\end{eqnarray}
where $v_{\gamma\pi}^B$ is the background transition potential and
$v_{\gamma\pi}^{\Delta}(E)$ corresponds to the contribution of the
bare $\Delta$ excitation. The resulting t-matrix can be decomposed
into two terms \cite{KY99}
\begin{eqnarray}
t_{\gamma\pi}(E)=t_{\gamma\pi}^B(E) +
t_{\gamma\pi}^{\Delta}(E),\label{eq:tgammapi33}
\end{eqnarray}
where
\begin{eqnarray}
t_{\gamma\pi}^B(E)=v_{\gamma\pi}^B+v_{\gamma\pi}^B\,g_0(E)\,t_{\pi
N}(E), \\
t_{\gamma\pi}^\Delta(E)=v_{\gamma\pi}^\Delta+v_{\gamma\pi}^\Delta\,g_0(E)
\,t_{\pi N}(E).
\end{eqnarray}
Here $t_{\gamma\pi}^B$ includes the contributions from the
nonresonant background  and renormalization of the  vertex
$\gamma^*N\Delta$. The advantage of such a decomposition is that
all the processes which start with the excitation of the bare
$\Delta$  are summed up in $t_{\gamma\pi}^\Delta$. Note that the
multipole decomposition of both $t_{\gamma\pi}^B$ and
$t_{\gamma\pi}^\Delta$ would take the same form as Eq.
(\ref{eq:backgr}).

As in MAID \cite{MAID}, the background  potential
$v_{\gamma\pi}^{B,\alpha}(W,Q^2)$ was described by Born terms
obtained  with an energy dependent mixing of
pseudovector-pseudoscalar $\pi NN$ coupling and t-channel vector
meson exchanges. The mixing parameters and coupling constants were
determined from an analysis of nonresonant multipoles in the
appropriate energy regions. In the new version of MAID, the $S$,
$P$, $D$ and $F$ waves of the background contributions are
unitarized  in accordance with the K-matrix approximation,
\begin{equation}
 t_{\gamma\pi}^{B,\alpha}({\rm MAID})=
 \exp{(i\delta^{(\alpha)})}\,\cos{\delta^{(\alpha)}}
 v_{\gamma\pi}^{B,\alpha}(W,Q^2).
\label{eq:bg00}
\end{equation}
From Eqs. (\ref{eq:backgr}) and  (\ref{eq:bg00}), one finds that
the difference between the background terms of MAID and of the
dynamical model is that off-shell rescattering contributions
(principal value integral) are not included in MAID. With
$v_{\gamma\pi}^{B,\alpha}$ specified, $t_{\gamma\pi}^{B,\alpha}$
can be evaluated according to Eq. (\ref{eq:backgr}) with a model
for the reaction matrix $R_{\pi N}^\alpha$. This is done with a
meson-exchange pion-nucleon model we have constructed
\cite{Hung94} within Bethe-Salpeter formulation. To take into account
of the inelastic effects at the higher energies, we replace
$\exp{i(\delta^{(\alpha)})} \cos{\delta^{(\alpha)}} = \frac 12
[\exp{(2i\delta^{(\alpha)})} +1]$ in Eq. (\ref{eq:backgr}) by
$\frac 12 [\eta_{\alpha}\exp{(2i\delta^{(\alpha)})} +1]$, where
$\eta_{\alpha}$ is the inelasticity. In our actual calculations,
both the $\pi N$ phase shifts $\delta^{(\alpha)}$ and
 inelasticity parameters $\eta_{\alpha}$ are taken from the analysis
of the GWU group \cite{VPI97}.

Following  Ref.~\cite{MAID},  we assume a Breit-Wigner form for
the resonance contribution $t_{\gamma\pi}^{R,\alpha}(W,Q^2)$ to
the total multipole amplitude,
\begin{equation}
t_{\gamma\pi}^{R,\alpha}(W,Q^2)\,=\,{\bar{\cal
A}}_{\alpha}^R(Q^2)\, \frac{f_{\gamma R}(W)\Gamma_R\,M_R\,f_{\pi
R}(W)}{M_R^2-W^2-iM_R\Gamma_R} \,e^{i\phi}, \label{eq:BW}
\end{equation}
where $f_{\pi R}$ is the usual Breit-Wigner factor describing the
decay of a resonance $R$ with total width $\Gamma_{R}(W)$ and
physical mass $M_R$. The expressions for $f_{\gamma R}, \, f_{\pi
R}$ and $\Gamma_R$ are given in Ref.~\cite{MAID}. The phase
$\phi(W)$ in Eq. (\ref{eq:BW}) is introduced to adjust the phase
of the total multipole to  equal  the corresponding $\pi N$  phase
shift $\delta^{(\alpha)}$. Because  $\phi=0$ at resonance,
$W=M_R$, this phase does not affect the $Q^2$ dependence of the
$\gamma N R$ vertex.

\section{Results and Discussions}
\noindent
We now concentrate on the $\Delta(1232)$ resonance. In this case
the magnetic dipole $({\bar{\cal A}}_M^{\Delta})$, the electric
$({\bar{\cal A}}_E^{\Delta})$ and Coulomb $({\bar{\cal
A}}_S^{\Delta})$ quadrupole form factors are related to the
conventional electromagnetic helicity amplitudes $A^\Delta_{1/2}$,
$A^\Delta_{3/2}$ and $S^\Delta_{1/2}$ by
\begin{eqnarray}
{\bar{\cal A}_M^\Delta}(Q^2)&=&-\frac{1}{2}(A^\Delta_{1/2} +
\sqrt{3} A^\Delta_{3/2}),\,\\ {\bar{\cal
A}_E^\Delta}(Q^2)&=&\frac{1}{2} (-A^\Delta_{1/2} +
\frac{1}{\sqrt{3}} A^\Delta_{3/2}),\,\\ {\bar{\cal
A}_S^\Delta}(Q^2)&=&-\frac{S^\Delta_{1/2}}{\sqrt{2}}\,.
\end{eqnarray}
We stress
that the physical meaning of these resonant amplitudes in
different models is different \cite{SL96,KY99}. In MAID, they
contain contributions from the background excitation and describe
the so called "dressed" $\gamma N\Delta$ vertex. However, in the
dynamical model the background excitation is included in
$t_{\gamma\pi}^{B,\alpha}$ and the electromagnetic vertex
${\bar{\cal A}}_{\alpha}^\Delta(Q^2)$ corresponds to the "bare"
vertex.

We further write, for electric ($\alpha=E$), magnetic ($\alpha=M$)
and Coulomb ($\alpha=S$) multipoles,
\begin{eqnarray}
{\bar{\cal A}}_{\alpha}^{\Delta}(Q^2)=X_{\alpha}^{\Delta}(Q^2)\,{\bar{\cal
A}}_{\alpha}^{\Delta}(0) \frac{ k}{k_W}\,F(Q^2),
\end{eqnarray}
where  $k_W = (W^2 - m_N^2)/2W$, $
k^2=Q^2+((W^2-m_N^2-Q^2)/2W)^2$. The form factor $F$ is taken to
be $ F(Q^2)=(1+\beta\,Q^2)\,e^{-\gamma Q^2}\,G_D(Q^2),$ where
$G_D(Q^2)=1/(1+Q^2/0.71)^2$ is the usual dipole form factor. The
parameters $\beta$ and $\gamma$ were determined by fitting
${\bar{\cal A}}_{M}^{\Delta}(Q^2)$ to the data for $G_M^*$ defined
by \cite{MAID,KY99,Ash}, $M_{1+}^{(3/2)}(M_{\Delta},Q^2)=(k/m_N)
\sqrt{3\alpha_e/8\Gamma_{exp} q_{\Delta}}\, G_M^*(Q^2),$ where
$\alpha_e=1/137$, $\Gamma_{exp}=115$ MeV, and $q_{\Delta}$ is the
pion momentum at $W=M_\Delta$. The values of ${\bar{\cal
A}}_M^{\Delta}(0)$ and ${\bar{\cal A}}_E^{\Delta}(0)$ were
determined by fitting to the multipoles obtained in the recent
analyses of the Mainz \cite{HDT} and GWU \cite{VPI97} groups. Both
$X_E$ and $X_S$ are to be determined by the experiment with
$X^\Delta_{\alpha}(0)=1$.

\subsection{Pion Photoproduction ($Q^2 = 0$)}
With background $t^B_{\gamma\pi}$, and the resonance contributions
associated with $\Delta(1232)$ and other resonances determined, we
obtain excellent agreement with the existing pion photoproduction
data, including cross sections and polarization data, from
threshold \cite{Kamalov01a} up to $1\, GeV$ photon lab energy
\cite{DMTweb}. Our results for $M_{1+}^{(3/2)}$ and
$E_{1+}^{(3/2)}$ multipoles at $Q^2=0$ are shown in Fig. 1 by
solid curves. The dashed curves denote the contribution from
$t_{\gamma\pi}^B$ only. The dotted curves represented the K-matrix
approximation to $t_{\gamma\pi}^B$, namely, without the principal
value integral term included.

\begin{figure}[h]
\vspace*{13pt} \centerline{\epsfig{file=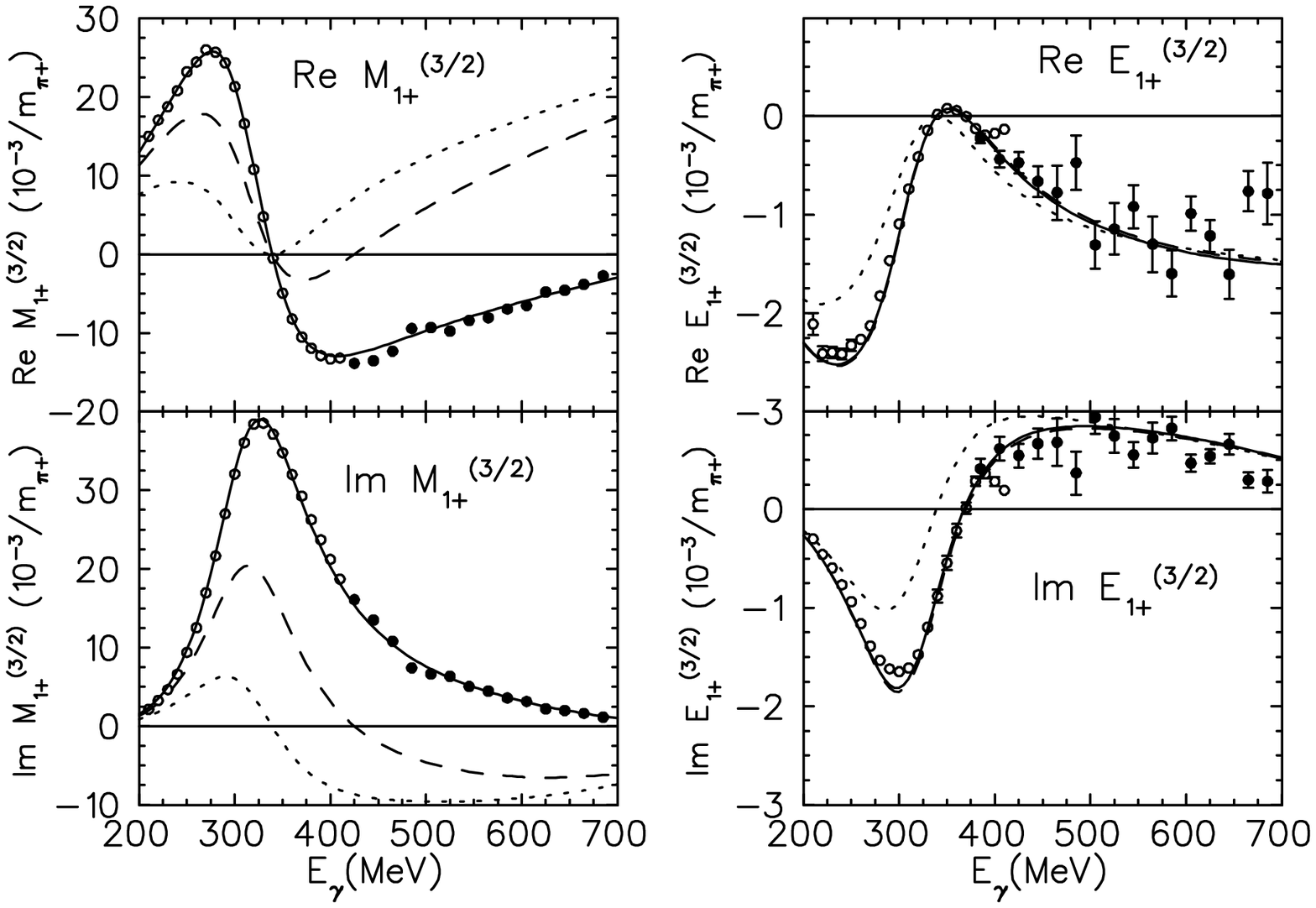,width=4.0in}}
\vspace*{13pt} \fcaption{ Real and imaginary parts of the
$M_{1+}^{(3/2)}$ and $E_{1+}^{(3/2)}$ multipoles. Dotted and
dashed curves are the results for the $t_{\gamma\pi}^B$ obtained
without and with principal value integral contribution in Eq. (4),
respectively. Solid curves are the full results with bare $\Delta$
excitation. For the $E_{1+}$ dashed and solid curves are
practically the same due to the small value of the bare
${\bar{\cal A}}^{\Delta}_E$. The open and full circles are the
results from the Mainz dispersion relation
analysis~\protect\cite{HDT} and  from the VPI
analysis~\protect\cite{VPI97}, respectively. }\label{figks2}
\end{figure}

The numerical values obtained for ${\bar{\cal A}}^\Delta_M$ and
${\bar{\cal A}}^\Delta_E$ , the helicity amplitudes, and
$\mu_{N\rightarrow\Delta}$ and $Q_{N\rightarrow\Delta}$   at $Q^2
=0$, are given in Table 1 along with the corresponding  "dressed"
values. At the resonance position $t_{\gamma\pi}^B$ vanishes
within K-matrix approximation and only principal value integral
term survives. The latter corresponds to the contribution where
$\Delta$ is excited by the pion produced via $v_{\gamma\pi}^B$.
Consequently, the addition of this contribution to
$t_{\gamma\pi}^{\Delta}$ can be considered as a dressing of the
$\gamma N \Delta$ vertex. The dressed helicity amplitudes obtained
in this way are in very good agreements  with PDG values.


\begin{table}[htbp]
\tcaption{Comparison of the "bare" and "dressed" values for the
amplitudes ${\bar{\cal A}}^{\Delta},\,$ ($10^{-3}\,GeV^{-1/2}$),
$\mu_{N\rightarrow\Delta}$ ($\mu_N$), and $Q_{N\rightarrow\Delta}$
($fm^2$) as compared with PDG values.}
\centerline{\footnotesize\smalllineskip
\begin{tabular}{|c|cccccc|}
\hline
 \hline
                    & ${\bar{\cal A}}^{\Delta}_M$  & ${\bar{\cal A}}^{\Delta}_E$ & $ A_{1/2}^{\Delta}$ & $ A_{3/2}^{\Delta}$  & $\mu_{N\rightarrow\Delta}$ & $Q_{N\rightarrow\Delta}$\\
\hline "bare"       & $ 158\pm 2 $                 & $ 0.4\pm 0.3 $              & $ -80 \pm 2 $       & $ -136 \pm 3 $       &  1.922                     & 0.009 \\
       "dressed"    & $ 289\pm 2 $                 & $-7\pm 0.4   $              & $ -134\pm 2$        & $ -256\pm 2  $       &  3.516                     &-0.081 \\
        PDG         & $ 293\pm 8 $                 & $-4.5 \pm 4.2 $             & $-140\pm 5 $        & $ -258\pm 6  $       &  3.512                     &-0.072 \\
\hline
\hline
\end{tabular}}
\end{table}

One notices that the bare values for the magnetic helicity
amplitudes determined above, which amount to only about $60\%$ of
the corresponding dressed values, are close to the predictions of
the constituent quark model (CQM).  The large reduction of the
helicity amplitudes from the dressed to the bares ones result from
the fact that the principal value integral part of Eq.
(\ref{eq:backgr}), which represents the effects of the off-shell
pion rescattering, contributes approximately for half of the
$M_{1+}$ as indicated by the dashed curves in Fig. 1.

For the standard Sach-type form factor
$G_M^{\Delta}(0)$~\cite{Jones}, which is proportional to $\mu_{N\rightarrow\Delta}(0)$, 
our bare and dressed values are
$1.65\pm 0.02$ and $3.06\pm 0.02$, respectively. On the other
hand, results of CQM calculations lie in the range 1.4--2.2
\cite{CQM}. From this result we conclude that pion rescattering is
the main mechanism responsible for the longstanding discrepancy in
the description of the magnetic $\gamma^*N\rightarrow\Delta$
transition within CQM. 

For $E_{1+}^{(3/2)}$, the dominance of
background and pion rescattering contributions further leads to a
very small bare values for electric  transition and $Q_{N\rightarrow\Delta}$. 
It implies that the bare $\Delta$ is almost spherical. We further note that the
dressed value for $Q_{N\rightarrow\Delta}$ is also small, negative but finite.
Since it is known \cite{Buchmann97,Dillon99,Blanpied01} that  $Q_{N\rightarrow\Delta}$ is proportional to
$Q_\Delta$, we conclude that the
physical $\Delta$ is oblate.


\subsection{Pion Electroproduction  ($Q^2 \ne 0$)}
\noindent The DMT model is used to analyze the recent JLab
differential cross section data \cite{Frolov99} on $p(e,e'p)\pi^0$
at high $Q^2$. All measured data, 751 points at $Q^2$= 2.8 and 867
points at $Q^2$= 4.0 (GeV/c)$^2$ covering the entire energy range
$1.1 < W < 1.4$ GeV, are included in our global fitting procedure.
We obtain a very good fit to the measured differential cross
sections \cite{Kamalov01}. In fact, the values of $\chi^2/d.o.f.$
for model are smaller \cite{Kamalov01} than  those obtained in
Ref. \cite{Frolov99}. Our results for the $G_M^*$ form factor are
shown in Fig. 2. Here the best fit is obtained with $\gamma=0.42$
(GeV/c)$^{-2}$ and $\beta=0.61$(GeV/c)$^{-2}$.

\begin{figure}[h]
\begin{center}
\epsfig{file=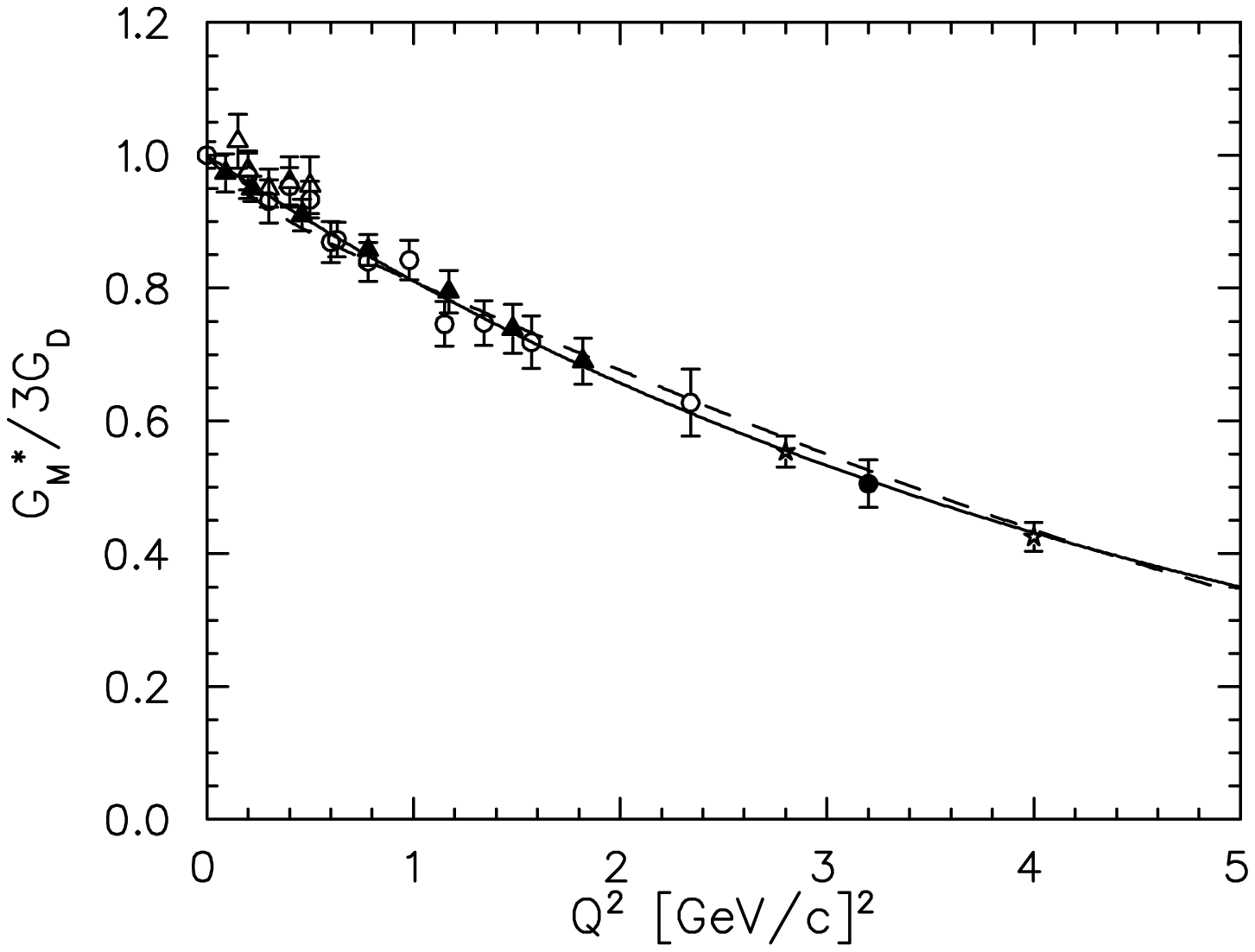, width=3.3in}
\end{center}
\fcaption{ The $Q^2$ dependence of the $G_M^*$ form factor. The
solid and dashed curves are the results of the MAID and DMT
 analyses, respectively. The data at $Q^2$=2.8 and 4.0
$(GeV/c)^2$ are from Ref.\protect \cite{Frolov99}, other data from
Refs.\protect \cite{EXPM}. } \label{fig2}
\end{figure}

With the resonance parameters $X_{\alpha}^{\Delta}(Q^2)$
determined from the fit, the ratios $R_{EM}$
and $ R_{SM}$ of the total multipoles and the
helicity amplitudes $A_{1/2}$ and $A_{3/2}$ can then be calculated
at resonance. We perform the calculations for both physical
$(p\pi^0)$ and isospin 3/2 channels and find them to agree with
each other. The extracted $Q^2$ dependence of the
$X_{\alpha}^{\Delta}$ parameters is: $X_{E}^{\Delta}=1+Q^4/2.4$, $
 X_{S}^{\Delta} = 1-10Q^2$, with $Q^2$ in units (GeV/c)$^2$.

Our extracted values for $R_{EM}$ and $R_{SM}$ and a comparison
with the results of Ref. \cite{Frolov99} are shown in Fig. 3. The main 
difference between our results and
those of Ref. \cite{Frolov99} is that our values of $R_{EM}$ show
a clear tendency to cross zero and change sign as $Q^2$ increases.
This is in contrast with the results obtained in the original
analysis \cite{Frolov99} of the data which concluded that $R_{EM}$
would stay negative and tend toward more negative
%
\begin{figure}[h]
\begin{center}
\epsfig{file=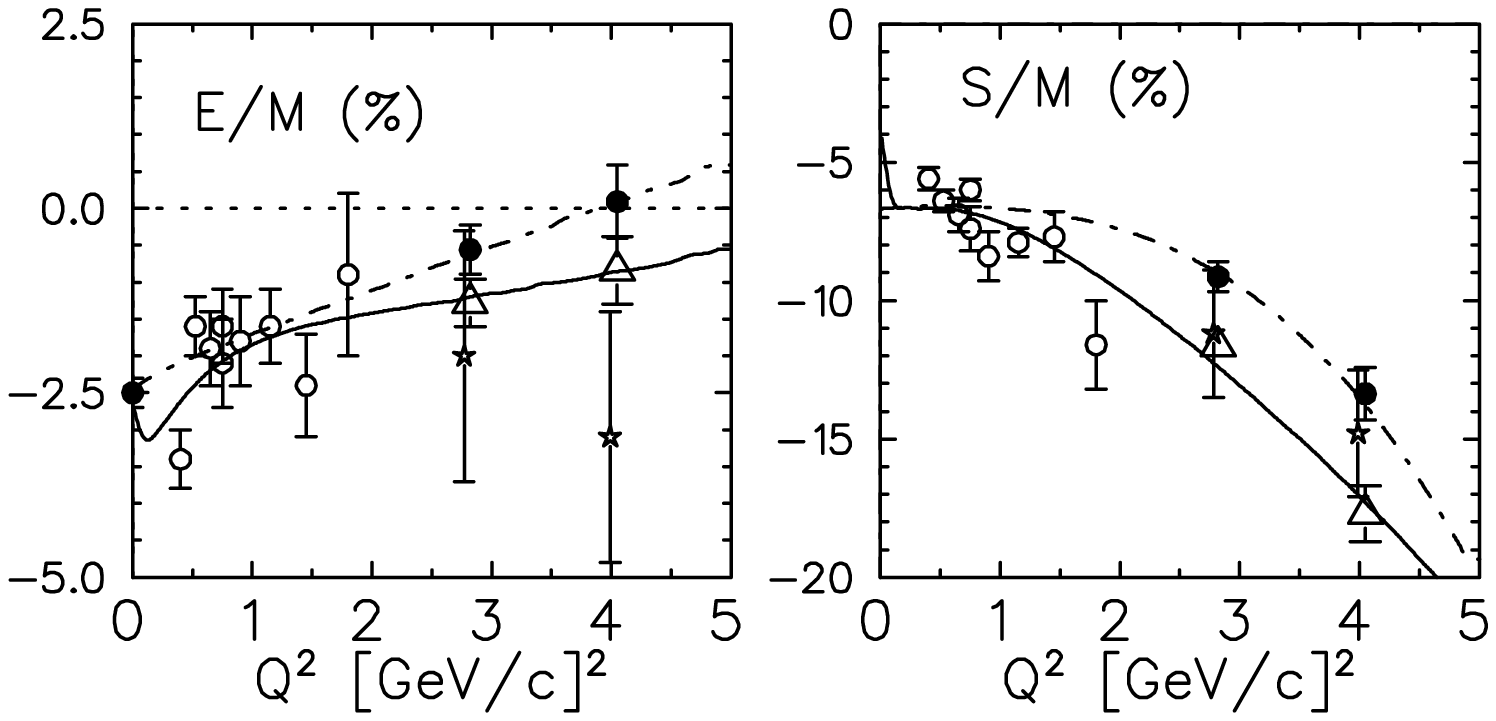, width=4.0in}
\end{center}
\fcaption{ The $Q^2$ dependence of the ratios $R_{EM}^{(p\pi^0)}$
and $R_{SM}^{(p\pi^0)}$ at $W=1232$ MeV. The solid and dash-dotted
curves are the DMT and MAID results, respectively. Results of
previous data analysis at $Q^2=0$ from Ref.\protect\cite{Beck97},
data at $Q^2$=2.8 and 4.0 $(GeV/c)^2$ from
Ref.\protect\cite{Frolov99} (stars). Results of our analysis at
$Q^2$=2.8 and 4.0 $(GeV/c)^2$  are obtained using MAID ($\bullet$)
and the dynamical models ($\bigtriangleup$). Open cycles are from
Ref.\protect\cite{Joo}.} \label{fig3}
\end{figure}
values with increasing $Q^2$. Furthermore, we find that
the absolute value of $R_{SM}$ is strongly increasing. Note that very recently
similar results were obtained by SAID group \cite{SAIDee}.

At low $Q^2$, the $Q^2$ evolution of both $R_{EM}$ and $R_{SM}$
obtained with DMT and MAID exhibits some marked difference, as can
be seen in Fig. 3. In particular, the value of $R_{SM}$ at $Q^2 =
0$ extracted with these two  models even differ by a factor of 2.
This is due to the fact that within MAID, the background
contribution of Eq. (\ref{eq:bg00}) vanishes at the resonance so
that $R_{EM}$ and $R_{SM}$ become  the ratios of the dressed form
factors  ${\bar{\cal A}}_{\alpha}^{\Delta}$. Therefore, if we
neglect the small influence of the $X_{\alpha}^{\Delta}(Q^2)$
factor at small $Q^2$, this leads to a rather smooth $Q^2$
dependence for the $R_{EM}$ and $R_{SM}$. In the DMT model,
both $E_{1^+}^{(3/2)}$ and $S_{1^+}^{(3/2)}$ are dominated by the
contribution from pion cloud \cite{KY99}, namely, the principal
value integral term in Eq. (\ref{eq:backgr}). Our results indicate
that the $Q^2$ dependence of the pion cloud contribution produces
negative slop at small $Q^2$. It is interesting to observe that
the recent calculation  of the two-body current contribution,
which in part includes the pion cloud effect, to the $R_{SM}$
within a constituent quark model \cite{buchmann01} also gives
results for $R_{SM}$ similar to our DMT values at small $Q^2$.
Similar effects were observed also in Refs.~\cite{ChPT,SL2001}.

In terms of helicity amplitudes, our results for a small $R_{EM}$
can be understood in that the extracted $A_{3/2}$ remains as large
as the helicity conserving $A_{1/2}$  up to $Q^2 = 4.0 (GeV/c)^2$,
as seen in Fig. 4, resulting in a small $E_{1^+}$. The
contributions from the bare $\Delta$ and pion cloud obtained with
DMT are also shown by the dashed and dotted curves, respectively.
Note that the latter drop faster than the bare $\Delta$
contribution. 
\begin{figure}[h]
\begin{center}
\epsfig{file=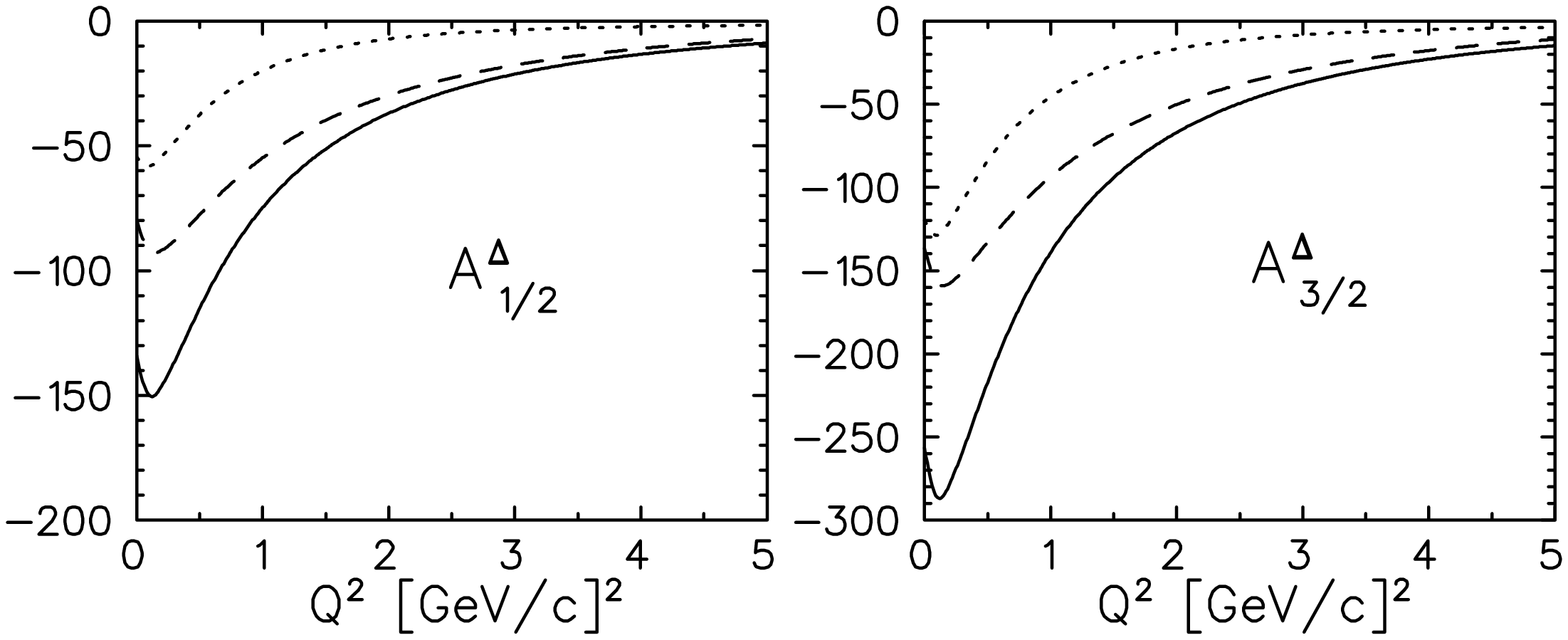,angle=90,width=4.0in}
\end{center}
\fcaption{ The $Q^2$ dependence of the bare (dashed curves) and
dressed (solid curves) helicity amplitudes $A_{1/2}$ and $A_{3/2}$
(in units 10$^{-3}$ GeV$^{-1/2}$) extracted with DMT. The dotted
curves are the pion cloud contributions.} \label{fig4}
\end{figure}
\begin{figure}[h]
\begin{center}
\epsfig{file=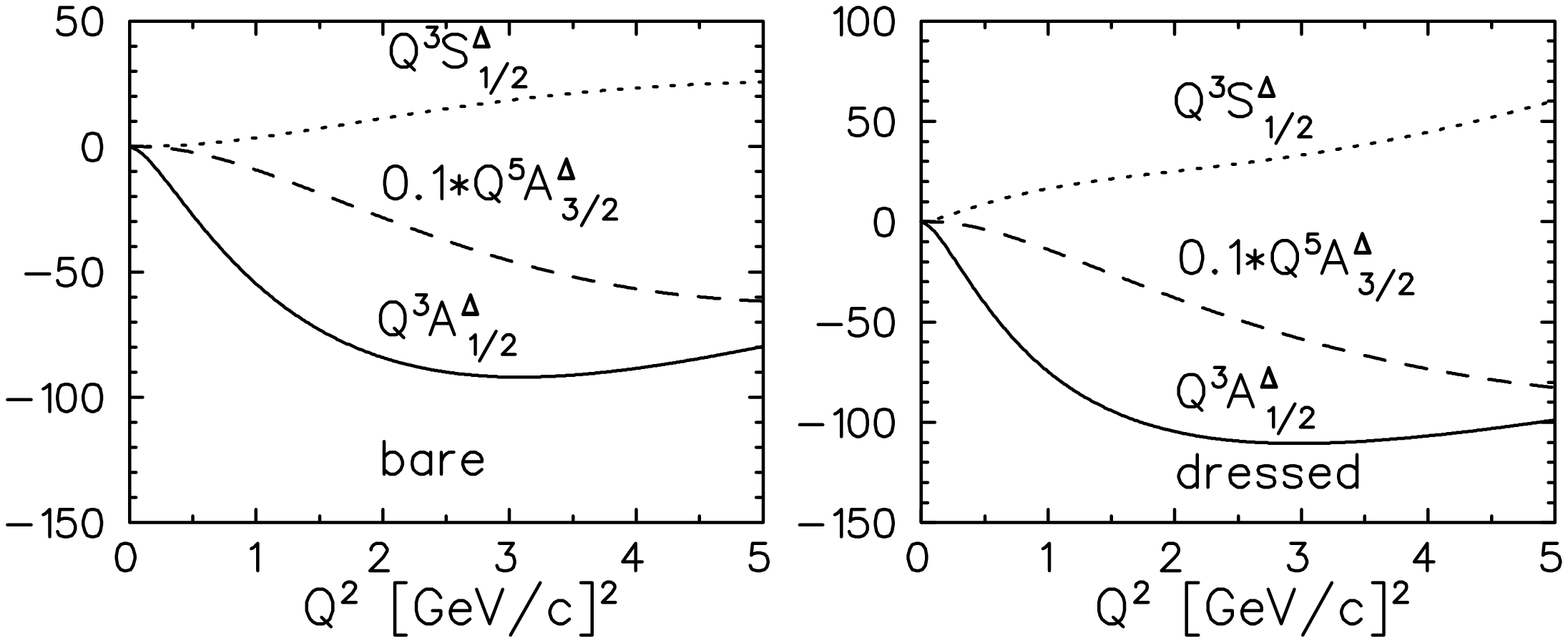, angle=90, width=4.0in}
\end{center}
\fcaption{ The $Q^2$ dependence of the  $Q^3 A_{1/2}^{\Delta}$
(solid curve) $Q^5 A_{3/2}^{\Delta}$ (dashed curve) and $Q^3
S_{1/2}^{\Delta}$ (dotted curve) amplitudes (in units 10$^{-3}$
GeV$^{n/2}$) obtained with DMT.} \label{fig5}
\end{figure}

Finally, we present in Fig. 5 our DMT results for
$Q^3A_{1/2}^{\Delta}, Q^5A_{3/2}^{\Delta},$ and
$Q^3S_{1/2}^{\Delta}$ to check the scaling behaviour of the bare
and dressed helicity amplitudes. Note that the scaling behavior
predicted by pQCD arises from the 3 quark (3q) Fock states in the
nucleon and $\Delta$ and should apply primarily to the bare
amplitudes. We find that the bare $S_{1/2}^{\Delta}$ and
$A_{1/2}^{\Delta}$ clearly starts exhibiting the pQCD scaling
behavior at about $Q^2 \ge 2.5 (GeV/c)^2$. However, it is
difficult to draw any definite conclusion for
$Q^5A_{3/2}^{\Delta}$. The dressed Coulomb form factor
$S_{1/2}^{\Delta}$ does not exhibit pQCD scaling behavior in the
considered $Q^2$-range. This is due to the fact that in this case
the dominant pion cloud contribution does not drop as fast as in
the transverse amplitudes. From these results, it appears likely
that scaling will set in earlier than the helicity conservation.
This is not surprising in the sense that the pQCD scaling behavior
is predicted based on the argument that, in exclusive reactions,
when the photon finds the nucleon in a small $3q$ Fock substate,
with dimensions comparable to the photon wavelength, then
processes with only two hard gluon exchanges dominate
\cite{Stoler}. On the other hand, hadron helicity would be
conserved only if this small $3q$ Fock state would further have a
spherically symmetric distribution amplitude such that $L_z=0$ and
the hadron helicity is the sum of individual quark helicities.

\section{Summary}
\noindent
In summary, we have studied the $\Delta$ structure by re-analyzing the recent JLab data for
electroproduction of the $\Delta(1232)$ resonance via
$p(e,e'p)\pi^0$ with  the DMT dynamical model for pion electroproduction, 
which give excellent descriptions of the existing data. 
Our results indicate that the bare
$\Delta$ is almost spherical and hence very difficult to be
directly excited via electric E2 and Coulomb C2 quardrupole
excitations. The experimentally observed $E_{1+}^{(3/2)}$ and
$S_{1+}^{(3/2)}$ multipoles are, to a very large extent, saturated
by the contribution from pion cloud, i.e., pion rescattering
effects. The negative value for the "dressed" value of $Q_{N\rightarrow\Delta}$
can be interpreted that the physical $\Delta$ is oblate
We find
that $A_{3/2}^{\Delta}$ is still as large as $A_{1/2}^{\Delta}$ at
$Q^2=4$ (GeV/c)$^2$, which implies that hadronic helicity
conservation is not yet observed in this region of $Q^2$.
Accordingly, our extracted values for $R_{EM}$ are still far from
the pQCD predicted value of $+100\%$. However, in contrast to
previous results we find that $R_{EM}$, starting from a small and
negative value at the real photon point, actually exhibits a clear
tendency to cross zero and change sign as $Q^2$ increases, while
the absolute value of $R_{SM}$ is strongly increasing. In regard
to the scaling, our analysis indicates that bare
$S_{1/2}^{\Delta}$ and $A_{1/2}^{\Delta}$, but not
$A_{3/2}^{\Delta}$, starts exhibiting the pQCD scaling behavior at
about $Q^2 \ge 2.5 (GeV/c)^2$. It appears likely that the onset of
scaling behavior might take place at a lower momentum transfer
than that of hadron helicity conservation.

\nonumsection{Acknowledgments}
\noindent
Part of the results presented here were obtained in collaboration with O. Hanstein,
D. Drechsel, and L. Tiator. This work is supported in part by the National
Science Council of the Republic of China under grant No. NSC 91-2112-M002-023. 

\nonumsection{References}

\end{document}